# Comparative Analysis of Packages and Algorithms for the Analysis of Spatially Resolved Transcriptomics Data

**Natalie Charitakis**[1], **Mirana Ramialison**[1,2,3,*] and **Hieu T. Nim**[1,2,3,*]

[1]Murdoch Children's Research Institute, Parkville, VIC, Australia

[2]Australian Regenerative Medicine Institute, Monash University, Clayton, VIC, Australia

[3]Systems Biology Institute, Clayton, VIC, Australia

Correspondence to:

natalie.charitakis@mcri.edu.au

mirana.ramialison@mcri.edu.au

hieu.nim@mcri.edu.au

*co-senior authors

**Abstract**

The technology to generate Spatially Resolved Transcriptomics (SRT) data is rapidly being improved and applied to investigate a variety of biological tissues. The ability to interrogate how spatially localised gene expression can lend new insight to different tissue development is critical, but the appropriate tools to analyse this data are still emerging. This chapter reviews available packages and pipelines for the analysis of different SRT datasets with a focus on identifying spatially variable genes (SVGs) alongside other aims, while discussing the importance of and challenges in establishing a standardised 'ground truth' in the biological data for benchmarking.

## Introduction

Despite the natural stochasticity that can disrupt biological processes such as organ development, biological systems consistently produce the same gene expression pattern with sufficient robustness such that the embryo forms correctly (nearly) every time. Furthermore, the genes typically work together in networks, requiring a systems-wide transcriptomic approaches to fully understand the spatial expression patterns. Many of these create well defined regions of cells within developing tissues that can be easily reproduced, demonstrating how the spatial location of the gene regulatory networks is critical for proper formation of tissues[1]. Determining these networks is an active study area in the emerging field of "spatial biology", and calls for specialised computational techniques, many of which have been developed very recently.

The merits and limitations of single cell RNA Sequencing (scRNA-Seq) have been well established[2,3] and the method successfully applied across varying organs and conditions[4–10]. scRNA-Seq is capable of identifying rare cell populations, including in disease states and developmental stages, however the method yields noisy, variable data with lots of technical variation[11]. Despite scRNA-Seq allowing the study of cellular heterogeneity and cell type hierarchy, the loss of spatial information prevents the systematic study of physiological structure/function relationships in various tissues and organs. This was part of the drive in the development of spatial transcriptomics (ST)[12] (now commercialised by 10x Genomics under the name Visium) and other spatially resolved transcriptomics (SRT) methods. The spatially-resolved gene expression pattern within the context of a tissue is critical to achieving full understanding of disease states and tissue development and function and the ability to investigate this is achievable using SRT[13].

Spatial transcriptomics is an area that is becoming more widely used and will continue to expand in the upcoming years[12]. Having been featured as Nature's Method of the Year in 2020, the technology and the analytical opportunities it provides are going keep growing rapidly[12]. As demonstrated by Figure 1, the number of papers published on spatial transcriptomics has greatly increased since 2016 when the first technology named 'spatial transcriptomics' was published[13]. Offering unprecedented spatial context to transcriptomic data it presents an invaluable tool for studying tissues and their cellular composition. As early as 2017, the merits of applying SRT to the discovery of spatial organisation of gene expression to improve transcriptional classification of cell types and localisation within a tissue had been discussed and even put to the test[14,15]. The potential applications of this technology are continuously improving and expanding, as demonstrated by the integration of different methods to improve the resolution of current SRT methods[16]. The different techniques available to generate SRT data and their merits have been discussed[14,17–19] a review of data analysis tools is as of yet lacking. With an emphasis on obtaining spatially resolved data sets with single cell resolution[12], the method,  aims and approaches to integrate and analyse the data generated still in flux, with a clear 'gold standard' yet to distinguish itself. This chapter will discuss some of the current packages and pipelines available to perform this analysis (Table 1).

## Methods for Downstream Analysis of Spatially Resolved Transcriptomics Data

As identifying the spatial expression patterns of genes and how they vary across tissue is a critical aim of spatial transcriptomics, many purpose-built tools for analysis of this data aim to identify spatially variable genes (SVGs) (Box 1)[1]. Building on the concept of highly variable genes in scRNA-Seq analysis, SVGs have a pattern of expression that depends on their location in the tissue and can give insight into biological function[20]. A complication of analysing these spatial transcriptomics data sets is accurately accounting for the spatial correlation across samples[21], and different methods can be employed to tackle this problem.  Various packages have been developed in primarily R or Python and are currently available to identify SVGs in spatial transcriptomic data sets.

<u>Identifying Spatially Variable Genes</u>

Among these, SpatialDE is a popular package based on Gaussian process (GP) regression which can clearly identify localized gene expression patterns for data sets containing temporal and/or spatial annotations[20]. SpatialDE can recognise SVGs by creating a model with two different terms reflecting distinct variance present in the data set. The first term captures the non-spatial variance present within the data while the second aims to capture the spatially related variance of gene expression within the data set; with the assumption that the covariance between cell's gene expression profile decreases with an increase in distance between the cells[20]. A ratio calculated using these terms can then be used as a measure of the level of gene expression variance attributable to spatial[20]. These are the key parameters used to fit the Gaussian model in a computationally efficient manner[20]. Testing to prove whether statistically significant SVGs are present is performed by comparing this model to a second one which lacks the spatial covariance parameter that represents a data set in which spatial localisation has no effect on gene expression patterns[20]. This process is repeated for each gene, and after correcting for multiple testing, the SVGs can be pulled out of the data set[20]. SpatialDE has the capability of taking this a step further by creating models with different covariance functions for SVGs and comparing them, this is in addition to the initial 10 Gaussian kernels it tests before selecting that with the lowest p-value. This creates the ability to determine whether each SVGs is most accurately expressed as a linear, periodic or general expression model[20]. However, for the data to fit certain underlying assumptions of the Gaussian model, two normalisation steps are performed, the first being a variance stabilising transformation[20,22]. It may affect the packages performance as the assumptions underlying the model and the necessary data transformations do not truly reflect the nature of the data[21]. A further functionality of SpatialDE is that it can implement an unsupervised learning technique built on the Gaussian Mixture Model to apply automatic expression histology (AEH) which can group together SVGs by their spatial expression pattern using hidden patterns learnt from the data[20]. The observation that SpatialDE may introduce false positives by labelling genes with low levels of expression as SVGs is an area which requires further investigation and can be improved upon in future releases of the package[22].

A package with the same goal as SpatialDE is SPARK (Spatial Pattern Recognition via Kernels) that employs a generalised linear spatial model (GLSM) with different spatial kernels to identify SVGs[22]. This model was built on previous work to take into consideration the effects of spatial correlation and covariate measurement error; it was built and tested on 2D data however it is capable of being expanded to 3D data sets[22]. As in the case of SpatialDE, SPARK models gene expression for each gene across all spatial coordinates, however this model operates under the assumption that the spatial data is non-Gaussian[22]. SPARK builds on other GLSMs by developing a hypothesis testing framework for the model[22]. The power of this hypothesis testing is linked to how the spatial kernel function accurately represents the spatial pattern of the gene represented in the model; and as different gene expression patterns will most accurately be represented by different spatial kernel functions, SPARK considers 10 different kernels (similarly to SpatialDE) based on commonly observed biological patterns[22]. Due to the heuristic nature of these kernels, this process could introduce biases that lead that package to choose more commonly observed biological patterns. SPARK can work with large data sets as it employs a penalised quasi-likelihood (PQL) algorithm for parameter estimation to circumvent the problem of the difficulty in solving GLSMs in short periods of time; this algorithm informs the parameters used in each of the spatial kernel functions. It further improves on the packages available at time of publication, SpatialDE and Trendsceek, by not performing a normalisation step on the data which decreases the power of the analysis[22]. A drawback of SpatialDE that SPARK corrects for, is to control for type 1 errors through the Cauchy combination rule thus giving it additional power when identifying SVGs[22]. The Cauchy combination rule groups the p-values generated from each spatial kernel function into a single p value while still controlling for type 1 errors which results in a single p-value per gene[22]. The final steps involve controlling for FDR across all p-values and then determining which are SVGs[22]. While SpatialDE and SPARK share the use of parametric test statistics there are a few critical differences between the packages[22]. As previously mentioned, SPARK does not model normalised data, while SpatialDE can only approximate p-values, SpatialDE calculates an exact p-value and once it obtains the initial set of statistically significant genes performs additional analysis to determine their p-values[22]. Furthermore, when validated against multiple data sets it performed just as well or better than SpatialDE and Trendsceek (described in next paragraph)[22]. When its ability to calculate true positives in two simulated data sets was tested across a total of six different spatial expression patterns with varying FDRs, SPARK outperformed Trendsceek and had better results than SpatialDE[22]. While with certain simulated data sets, SPARK and Trendsceek performed similarly in computing well calibrated p-values, SpatialDE's overlooked assigning certain true positives[22]. While the SPARK paper only tests the package's performance against SpatialDE and Trendsceek, it outperformed both in terms of number of SVGs identified when validated against a spatial transcriptomics mouse olfactory bulb data set[22]. However not all genes identified by SpatialDE overlapped with those identified by SPARK[22]. Despite this, the newly identified SVGs are in line with markers specific to the tissue they were annotated in and GO enrichment analysis adds further confidence that the majority of these newly identified SVGs are biologically relevant[22]. In terms of computational efficiency, when running with 10 parallel CPU threads, SPARK was more computationally efficient than the same analysis run on a single-threaded SpatialDE (although the difference in this instance is minimal) and Trendsceek, its single threaded performance is consistently less efficient than SpatialDE across 4 datasets of varying sizes[22].

Trendsceek is one of the earlier packages developed to identify SVGs using a nonparametric approach[23]. Trendsceek individually assesses each gene and normalises its expression through a log10 transformation[23]. It relies on a marked point process to model gene expression and cell location and later will test the null hypothesis by generating four non-parametric test statistics[23]. These four test statistics yield four p-values, and genes with a minimum of 1 p-value $\leq 0.05$ after adjustment for multiple testing using the Benjamini-Hochberg method is determined to be an SVG[23]. A key difference that separates Trendsceek from SpatialDE and SPARK is its computing of non-parametric test statistics meaning it lacks an underlying generative model. Trendsceek was tested against simulated data sets and it demonstrated very low power to identify SVGs when they were present if less than 5% of cells in the data set had varying levels of expression[23]. This implies that as SRT datasets continue to increase in size, Trendsceek will not be able to distinguish SVGs present in a very small subset of cells within a tissue. When Trendsceek's performance in identifying SVGs across two spatial transcriptomics data sets is compared to SpatialDE and SPARK, it identified fewer SVGs, with numbers almost 10 times lower than the other packages[22]. When compared to different packages in other studies, Trendscreek struggled to identify SVGs in real datasets, while other packages were able to.

Each new package developed aims to address the shortcomings of those already published, for example BOOST-GP claims that many popular substitutes such as SpatialDE, SPARK and Trendsceek, do not account for the substantial proportion of zero counts present in the data set and the effect this can have analysis[21]. Therefore BOOST-GP puts forth a new Bayesian hierarchical model aimed at accounting for the considerable number of 0 counts present in spatial data sets, that other packages published up to this point had neglected[21]. A key difference to other packages is that BOOST-GP employs a negative binomial distribution when modelling count data which should account for its observed over-dispersion[21]. This is more like the methods used by popular bulk RNA-Seq analysis packages rather than other spatial transcriptomics packages explored thus far[21]. BOOST-GP's performance was compared to that of SpatialDE's, SPARK and Trendsceek when there were false zeros present in the data and BOOST-GP was clearly the most adept at handling this complication, even if it still presented significant difficulties in retrieving a good Matthews correlation coefficient (used to determine the tool's accuracy) on a synthetic data set[21]. Furthermore, depending on the spatial pattern of expression of the gene, the accuracy of BOOST-GP can differ slightly[21]. Alternatively, when the tool was tested on two real data sets it was found that SPARK identified more SVGs than BOOST-GP, however, SpatialDE discovered the least[21]. In the analysis of human breast cancer data, despite identifying fewer SVGs than SPARK, BOOST-GP was able to identify novel, biologically relevant terms in the GO analysis adding to its value in the analysis of spatial transcriptomics data[21].

As larger datasets become increasingly common, packages must be created to efficiently analyse the vast amounts of data generated by spatial transcriptomics experiments. One of the newer packages is SOMDE[24]. Built in python, SOMDE aims to identify SVGs in large-scale datasets[24]. By using a self-organising map (SOM) neural network and a Gaussian process to model the data it can identify SVGs in large datasets much faster than SpatialDE, SPARK or Trendsceek[24]. This is achieved as the data is organised into different nodes by the SOM neural network, the location and expression data on the node level is used through the Gaussian process to identify the SVGs present in the data[24]. The organisation of data into nodes minimises the sample space while preserving the original spatial organisation and expression data[24]. The next stage which uses a Gaussian process identifies the SVGs from the reduced sample space[24]. As seen in packages such as SpatialDE and BOOST-GP, Gaussian process are a popular method for identifying SVGs[20,21]. SOMDE also uses a log ratio test like that employed by SpatialDE to test the statistical significance of the spatial expression variability of each gene[7]. When SOMDE was applied to discover the SVGs of five different data sets it was able to do so without significant increase in computational time as the size of the data set increased, yielding results in under 5 minutes for the largest data set with over 20,000 data sites[24]. It also demonstrated a faster running time compared to Giotto and SpatialDE on three differently sized data sets used for validation[24]. Despite this, the package lacks validation on a data set of single cell resolution[24]. When its performance was compared to scGCO and SpatialDE on a simulated data set, SOMDE consistently outperformed scGCO but only had an improved performance compared to SpatialDE when a high dropout rate is incorporated into the data set[24]. When its performance was compared to real data sets, most of the SVGs identified by SOMDE overlap with those identified by packages like scGCO, SPARK and SpatialDE[24].

Other methods have been developed to identify SVGs that differ from those presented thus far. One of these methods has been implemented in a python package called scGCO which employs graph cut algorithms to identify spatial genes[25]. scGCO first produces a graph by performing a Delaunay triangulation in which only true cell neighbours are connected by edges allowing an accurate representation of cellular interactions in a sparse graph which is not memory intensive[25]. Subsequently, Voronoi diagrams are created which have previously been used to model cells[25]. Using a Markov random field (MRF) model and adapting methods traditionally used in object identification in images, scGCO can classify cells into two categories which provides efficient, low polynomial time computing and a result which is globally optimal[25]. Much like SpatialDE, scGCO employs Gaussian Mixture modelling but uses it to classify each gene's expression to ensure more accurate classification of cell types based on their gene expression[20,25]. The performance of SVG was tested against a spatial transcriptomics data set obtained from a mouse olfactory bulb and compared to results obtained from the same data by SpatialDE[25]. A more comprehensive review of scGCO against different packages would be beneficial to obtain a holistic understanding of its improved performance in SVG detection. scGCO successfully identified over 1,000 additional SVGs compared to SpatialDE, and at an FDR cut-off of 0.01 rather than 0.05[25]. scGCO was able to identify the majority of SVGs also picked up by SpatialDE and the SVGs formed clear clusters each with its own spatial pattern. These results were consistent when validation was repeated across replicate mouse olfactory bulb data[25]. However, while scGCO yielded a smaller number of unreproducible SVGs across the different replicate data sets than SpatialDE, ~35% of identified SVGs were still unreproducible (an 11% reduction from SpatialDE)[25]. If replicate data sets are available for studies, then this is something that should be investigated further across all packages, resulting in the exclusion of non-reproducible SVGs for a more accurate final subset of SVGs. Additionally, when comparing between regions of the mouse olfactory bulb scGCO was more adept at identifying SVGs than SpatialDE, while neither method entirely recovered all marker genes reported in the study which published the data set[25]. Additional validation was performed using data from breast cancer biopsies, with scGCO having a similar improved performed compared to SpatialDE when employed on the mouse olfactory bulb data set. Furthermore, the SVGs identified by SpatialDE within the breast cancer data set did not maintain consistent clustering pattern[25]. scGCO's performance on other spatial transcriptomics data sets was equally as robusts[25]. scGCO also performed better in terms of computational time and memory required than SpatialDE and Trendsceek when used to analyse a simulated data set with up to a million cells.

Identifying Spatially Variable Genes and More

As evidenced by the packages reviewed so far, GPs are popular method for analysis spatial transcriptomics data as they can model its spatial dependence. To this end, as new packages are developed many are built on alternative GP regression models such as GPcounts[26].

GPcounts can be used to model either spatial or temporal, large-scale scRNA-Seq data through modelling count data using a negative binomial (NB) likelihood[26]. The NB likelihood model should more accurately capture the distribution of gene expression data compared to Gaussian likelihood model as it accounts for possible heteroscedastic noise and the presence of many zero-counts but requires UMI normalisation to be applied[26]. Furthermore, GPcounts evaluates its performance across different simulated data sets when it implements different underlying likelihood models to determine under which conditions each yields the best results[26]. Subsequently, it can be observed that employing a NB likelihood was effective in producing accurately identified SVGs in the package BOOST-GP[21]. However, GPcounts's primary aim is not to identify SVGs, it is also able to identify differential expressed genes, perform pseudotime inference and then identify branching genes and discover temporal trajectories, widening its scope compared to most packages[26]. The GP model is stochastic and non-parametric and there is a choice of kernel to find one that most accurately models the data, similarly to the step employed by SpatialDE[20], and this is determined by the Bayesian Inference Criterion[26]. Using SpatialDE as a benchmark, GPcounts builds on and alters many of the steps implemented by SpatialDE[26]. This applies from the testing procedures used to determine SVGs and differentially expressed genes p-values to the type of normalisation applied to the data[26]. GPcounts has also implemented the additional step of a built-in check during its kernel function hyperparameter estimation to minimise the problems of getting stuck in a local optimum by restarting the optimisation is this is suspected[26]. This is so far one of the only optimisation-based method that has implemented this kind of self-check and could give GP-counts a distinct advantage in the accurate identification of SVGs.  An improved assessment of GPcounts performance when detecting differentially expressed genes would be to evaluate the package on published data sets in addition to the simulated data[26]. When evaluated for its identification of SVGs, GPcounts did use a real mouse olfactory bulb data set and compared its performance to SpatialDE, SPARK and Trendsceek[26]. GPcounts identifies the most SVGs out of any of the packages, with the vast majority of identified SVGs at a 5% FDR overlapping with those identified by SpatialDE and SPARK[26]. The unique SVGs identified by GPcounts have spatial patterns that match those depicted in the Allen Brain Atlas, indicating a high confidence in these findings[26]. GPcounts also identified 90% of the biologically important marker genes expressed in the dataset, although SPARK had a similar performance as it identified 80%[26], while SpatialDE identified only 30% of the marker genes[26].

Certain frameworks have been developed with a particular SRT technology in mind, in combination with addressing an area of data analysis the developers deem lacking. One of these is the STUtility workflow created in R and based and built on the Seurat analysis tool[27]. Aiming to develop a package that allows the user to visualise multiple experiments in conjunction to create a 3D view of tissue, STUtility builds on well-established methods of analysis (moulded by those established for scRNA-Seq analysis) to focus on novel data visualization[27]. Highlighting the importance of data normalisation and transformation to deconvolute technical noise from meaningful biological insight, the package uses a regularized negative binomial regression model successfully implemented in Seurat for normalisation[27]. The image processing capabilities of STUtility focus on alignment, automatic or manual, of multiple samples in addition to the removal of background noise[27]. The removal of background noise – called masking in the study – is an integral part of image processing and allows the inside and outside of the tissue to be defined as well as decreasing the images' storage requirements[27]. To automatically align multiple samples the package identifies a reference image then uses an iterative closest point (ICP) algorithm to align the remaining samples to the reference, which can then be reconstructed into a 3D tissue model[27]. While this method of creating a 3D model is not one which yields the most precise cell segmentation, this trade-off yields greater computational efficiency and still gives a faithful reconstruction of tissue morphology[27].

Implementation of a k-means clustering algorithms allows the package to clearly define the boundaries of the tissue[27]. For the sequencing data, STUtility leans heavily on the functions created by the package Seurat[27]. A decomposition of the normalised gene data called non-negative matrix factorization (NMF) is used to choose gene drivers and create a low dimensional representation of the data to be used in defining clusters and nearest neighbours[27]. To obtain genes whose expression demonstrates spatial patterns a connection network is created for each spot which allows the package to calculate the spatial-lag of each gene across spots. This is one of the inputs – the other being the normalised counts – used to calculate spatial correlation across the sample[27]. It's ability to visualise spatial distinct features is clearly demonstrated and determining the spatial relation of gene expression to tissue areas (e.g., a tumour). STUtility is also able to identify SVGs using neighbourhood networks, but it's accuracy in performing this function is not compared to other packages[27]. Other capabilities were tested on a variety of human and mouse tissues[27]. For both mouse brain and human breast cancer tissue samples spatial gene expression patterns can be clearly identified[27]. STUtility allows for the manual alignment of multiple images, however a comparison as to the accuracy of this method compared to the automatic alignment is not offered and depending on the expertise of the user may vary significantly[27]. Furthermore, while its implementation of neighbourhood networks offers a promising method to define subsections within a tissue and the heterogeneity within, as would be beneficial during the study of tumours, to see how well this correlates to the actual tissues heterogeneity of the sample is not reported[27,28].

Assigning Lost Transcripts

Other packages have been developed with the aims of addressing gaps in analysis that have not been adequately accounted for; one such package is Sparcle[29]. When attempting to obtain an accurate gene counts matrix from image based spatial transcriptomics techniques, often many transcripts are not assigned to cells after segmentation is performed, leading to a loss of data[29].

Sparcle aims to recapture the data from these ‘dangling’ transcripts[29]. Developed to be used in conjunction with data from any smFISH technology, Sparcle can build a probabilistic model which allows assignment of these dangling transcripts to the appropriate neighbouring cells using a maximum likelihood estimation (MLE). The MLE considers the dangling mRNA’s distance to other transcripts, nearby cells and genes’ covariance when calculating which nearby cell the transcript should most accurately be assigned to[29]. Similarly to other packages, Sparcle assumes that the most accurate representation of gene expression can be modelled using a multivariate Gaussian distribution[20,29]. Sparcle can employ two clustering methods when it first groups the cells in the chosen Field of Vision (FOV) by cell type based on a global count matrix: DPMM and Phenograph. Phenograph is an algorithm developed to cluster cell phenotypes in high-dimensions single cell data and was originally applied to data from acute myeloid leukemia[30]. Dirichlet Process Mixture Models (DPMM) is a stochastic process which can feature all the individual Gaussian distributions for the expression of each gene and allows Sparcle to model all these distributions[31]. While having the additional flexibility to employ either of algorithm at the clustering step, during its validation Sparcle reports data based on the Phenograph algorithm but not on the performance when using DPMM, nor does it specify in which instance one method should be favoured over another[29]. When used to assign dangling transcripts to a MERFISH data set, Sparcle was able to assign 68% or almost 2 million missed transcripts, and validation with scRNA-Seq data confirmed that the proportion of cell types assigned post used of Sparcle more closely matched the scRNA-Seq data[29]. Validation against other neuronal data sets returned similarly desirable results. Despite this, there are limitations to the use of Sparcle. For example, when the programme draws an area around each dangling transcript that should mimic the size of a cell, the size of this area is optimised to the size of an average neuronal cell, meaning the package might not be well suited to non-neuronal data[29]. Sparcle can run on approximately 80 cells in under 10 minutes with impressive mRNA recovery over three iterations, however additional data on how this would scale with larger data sets is lacking, potentially causing computational bottlenecks in bigger data sets[29]. It claims to improve on packages that remove the cell segmentation step entirely, such as Baysor and SSAM, by removing the need for *a priori* knowledge of the data set and not assuming that the cellular mRNA can be modelled by a uniform distribution[29]. However, some further improvements could be made to enhance the performance such as staining cellular membranes to better understand the size of neighbouring cells rather than estimating based on an area around the nucleus and calculated an estimate of the prior distribution of a gene's localised transcripts[29].

Estimation of Cell Type Composition

Identifying SVGs was the primary focus of the initial packages developed, but it is important to note that packages with alternative aims are increasingly being published. For example, SpatialDWLS was created to improve the identification of different cell types at locations in the data sets which do not have single-cell resolution[32]. This is termed cell-type deconvolution[32]. Other published packages have been developed for this aim but SpatialDWLS claims to improve on the results of these packages[32]. How SpatialDWLS performs cell-type deconvolution can summarised in two steps, the first uses a cell-type enrichment analysis method to identify which kinds of cells have a high probability of being at each location and the second uses an extension of the dampened weighted least squares (DWLS) method to pin-point the precise composition of cell types at the specified location[32]. Firstly, signature genes can either be supplied by the user to identified by differential expression analysis[32]. Building on the previously developed DWLS method for scRNA-Seq data this was extended to spatial transcriptomics data by incorporating the signature genes step[32]. Furthermore, SpatialDWLS builds on clustering and gene marker identification used in Giotto[32,33]. This would imply that any shortcoming with Giotto's performance in these areas would be transferred to SpatialDWLS. When evaluated on a simulated spatial transcriptomics dataset, SpatialDWLS outperformed RCTD and stereoscope in terms of having a lower Root Mean Square Error (RMSE) and in terms of computational time[32]. However, when its performance was tested against a real mouse brain Visium data set, SpatialDWLS's performance was not benchmarked against the other three packages thus making its performance on real data unclear[32]. Despite this the authors reported that the spatial location of the cell types assigned by SpatialDWLS were consistent with those reported in the Allen Mouse Brain Atlas[32]. An interesting application of this package was to identify the change of cell type organisation in a spatial-temporal context throughout embryonic heart development[32]. In addition to quantifying an increase in ventricular cardiomyocytes and smooth muscle cells as time went on, by calculating the assortativity coefficient (here used as a measure of whether neighbouring cells were of the same type) the study was able to determine that spatial organisation of the developing heart becomes increasingly defined in terms of neighbourhoods of cell types during development[32].

Assigning cell types to a spatial transcriptomics dataset can be approached more than one way. By incorporating *a priori* knowledge to a probabilistic likelihood function, FICT (FISH Iterative Cell Type assignment) can blend expression and spatial information to assign cell type to spatial transcriptomics data sets[34]. This is achieved by creating a generative mixture model using a reduced dimensions representation of expression levels through a denoising autoencoder and assigning each cell as cell type defined by its neighbourhood (represented in an undirected graph); the parameters of this model can be learnt by an expectation maximization approach which is an iterative process[34]. Finally, the cell can be classified by posterior distribution of the model[34]. During this process, the problem of over reliance on expression data needs to be addressed, which occurs because in a dataset it is likely that there are more genes being expressed than cell types present[34]. To circumvent this problem, a named power factor acts as a weight term to balance the dimensionally reduced expression component with the spatial component[34]. The package was validated using three simulated and real data sets and compared it to the results of GMM, scanpy, Seurat and smfishHmrf[34]. Across all three simulated data sets, FICT has the highest median accuracy, reaching a high of approximately 0.89 in one of the simulated data sets[34]. When evaluated on a real MERFISH mouse hypothalamus data set, no ground truth as to the location of different cell types is available for this data clustering results between parameters learnt for the same models on different animals are compared by the Adjusted Rand Index. When comparing across this metric, FICT is more consistent in applying clusters to the majority of the paired animals, indicating its superior performance in assigning cell type clusters[34]. FICT as has the potential to identify novel subclusters within the data set[34]. However, FICT's performance drops when applied to data sets with smaller numbers of cells, although this is observed across all packages validated[34]. Furthermore, its decreased performance was still in line with packages with similar functions and as spatial transcriptomics data sets become larger this should not interfere with FICT being applied in future[16]. However, despite its greater accuracy when applied to larger datasets, FICT's runtime in these instances could still be improved[16].

RCTD is another package created with the final aim of identifying cell types in a spatial transcriptomics data set[35]. While identifying SVGs is extremely informative, it is important to understand how the role of underlying cell types contributes to gene's spatially variable expression patterns[35]. Robust Cell Type Decomposition (RCTD) makes use of annotated scRNA-Seq data to create cell type profiles for expected cell populations in the data, then can labels spatial transcriptomics pixels with cell types using a supervised learning method[35]. As one of major hurdle in this analysis is the fact that current spatial transcriptomics data sets can contain multiple cell types within a single pixel, RCTD can also fit a statistical model to determine multiple cell types present within a pixel and normalise across platform effects between the scRNA-Seq and SRT datasets[35]. To achieve this RCTD first creates a spatial map of cell types and estimates the number of different cell types in each pixel where the gene counts are assumed to have a Poisson distribution[35]. This should circumvent the problem introduced by current unsupervised learning methods that overlook clustering cells that co-localise transcriptionally as well as spatially[35]. Using this approach RCTD was able to classify cells across platforms with almost 90% accuracy. However, as with any supervised learning approach, the cell types one can detect using this tool are limited to how accurately and fully the reference data set is annotated, which may present difficulties. Also, while the study tested RCTD using references and data sets generated by many different kinds of scRNA-Seq and SRT technology, the effects that specific platforms may have on cell-type assignment is still undetermined.

Spot-by-Spot Clustering

A common step in the analysis of many kinds of omics data sets is to perform clustering, and this is prevalent when analysing SRT data. This section will discuss techniques that cluster spots on a SRT array, that may contain multiple cell types, based on the overall gene expression profile of the spot[36]. Despite being common, this is not a straightforward step. Understanding the results after different iterations can prove difficult, as does choosing the correct hyperparameters[36]. This is further confounded as each barcode is associated to multiple cells[36]. To address these issues, an R package called SpatialCPie was developed which focuses on clustering spots on the array based on the gene expression profile to allow annotation of regions of the tissue[36]. SpatialCPie allows the user to choose which algorithm to implement and clusters the data at different resolutions from the start[36]. The user is then free to choose which conformations of clusters created at which resolution most accurately represent their data. By creating a cluster graph and an array plot, SpatialCPie gives the user varied insight into how different resolution affects the clustering outcomes[36]. The cluster graph relates how the different clusters relate to one another across different resolutions, and as new clusters emerge at higher resolutions from which cluster they originate[36]. The edges of the graph link the percentage of spots in new clusters that descend from different lower resolution clusters[36]. The second visualisation method is the array plot which represents the ST array, but each spot is depicted as a pie cart that shows how similar the gene expression is between cluster centroids and spatial regions[36]. SpatialCPie offers the novel, to the best of the authors' knowledge, option to choose a particular region of the dataset for further sub-clustering which may be appropriate depending on the tissue of interest[36]. While SpatialCPie only compares itself to ST viewer - in a limited capacity - its overall performance is promising[36]. However, additional validation of its performance compared to other similar packages such as ST viewer would be beneficial to understand its accuracy.

Pipelines

As the area of spatial transcriptomics continues to expand, pipelines rather than just analysis packages will become more common place. One of the first available pipelines written in R is Giotto, which is a platform that can be used on both transcriptomics and proteomics data; it is divided into a data analysis and visualisation module[33]. With a focus on being user-friendly and reproducible, Giotto does provide the opportunity for more complex spatial analysis using HMRF models[33]. As a foundation, Giotto creates a neighbourhood network of cells and a spatial grid for downstream analysis which includes ligand-receptor identification, gene expression pattern analysis and determining preferential cell neighbours[33]. Giotto is tested on ten different data sets obtained with varying technologies and from varied tissues to examine its performance across a range of benchmarks[33]. The initial steps in the analysis are similar to those performed in scRNA-Seq analysis but Giotto does offer three different algorithms for identifying marker genes, one of which (Gini) was specifically developed for the pipeline, which differ in their strength in identifying particular kinds of marker genes[33]. The Scran method evaluates the markers between two groups of cells by running t-test (default) and then determining marker genes[37]. Mast identifies markers genes between two cell groups by employing a hurdle model[38]. The Gini algorithms scores markers genes within a cluster based on Gini coefficients which was developed to identify rare cell types from an adapted model implemented in the social sciences[39]. All of these algorithms were developed to score markers genes between clusters in single cell data sets. When evaluated, Gini discovered the most marker genes overall for the 12 cell types when compared to Mast and Scran, however when analysing the sensitivity and specificity for each method when identifying their top 20 markers Gini had the lowest sensitivity but highest specificity in both the endothelial and oligodendrocyte populations[33]. The sensitivity and specificity of each algorithm vary slightly across the different cell populations they investigated when evaluated against a sequential fluorescence in situ hybridization (seqFISH+) somatosensory cortex dataset and this is important when deciding which algorithm to employ, furthermore this needs to be tested against data sets generated from different biological material and technologies to best understand the true limitations of each algorithm[33]. Giotto also has analysis pipelines designed specifically for spatial transcriptomics data sets with lower resolution[33]. By using one of three algorithms to provide an enrichment score between a location's expression pattern and a cell's gene signature it is possible to assign a cell type to a location which contains more than one cell[33]. Once again, the availability of multiple algorithms at this step allows which require different inputs allows Giotto to be flexibly implemented to a number of different datasets[33]. These three enrichment algorithms were validated on a simulated dataset similar to one generated using seqFISH+ with the hypergeometric algorithms having the lowest AUC score (0.8) and both PAGE and RANK scoring similarly well when predicting cell type at a particular location[33]. When applied to real data sets, the two best scoring algorithms RANK and PAGE performed well and should be used when employing the Giotto pipeline[33]. To analyse spatial patterns of gene expression, Giotto creates a spatial network to represent the data using a Delaunay triangulation network, which is the same as the method employed by scGCO[25,33]. While the option is available to alternatively construct a spatial network with two different methods offering the user greater control on downstream parameters, the analysis results appear insensitive to these adjustments[33]. To uncover SVGs Giotto introduces two new methods, BinSpect-kmeans and BinSpect rank, as well as incorporated methods from SpatialDE, Trendsceek and SPARK[33]. When evaluated, each of the methods identified unique SVGs, with 103 genes being identified by all five methods[33].

As the field of SRT continues to expand so will the analytical tools available. As an increasing number of down-stream analysis packages are published for SVG identification amongst other analysis, pipeline and frameworks will become increasingly complex in the scope of their abilities. A new framework developed to combine and encompass all aspects of analysis for spatial -omics technology is Squidpy[28]. While not built specifically for the analysis of spatial transcriptomics data the Squidpy framework developed in Python brings common tools for analysis and visualisation to any spatial -omics data and takes advantage of the additional information available to improve exploration[28]. Offering a broader and more modular approach than Giotto, Squidpy offers the opportunity for other packages to be easily integrated into its pre-existing framework to expand its capabilities[28]. Squidpy will store the image data in an Image Container and create a neighbourhood graph of spatial coordinates so that it can be used on a wide array of technologies[28]. A feature of Squidpy that adds additional analytical opportunity is its in-built image analysis tools[28]. While the packages discussed so far require an image as part of the input for analysis, none extend so far as to allow the user to investigate the data contained in this image to the same extent as Squidpy, which is the capability that differentiates it most from Giotto[28]. The first step in the investigation of cellular neighbourhoods and spatial patterns is the construction of a spatial graph[28]. When compared to similar processes in Giotto, Squidpy had a more efficient run time when constructing both a spatial graph and calculating neighbourhood enrichment, although for data sets with a smaller number of observations the difference was not great[28]. Despite offering an interesting perspective on the direction of spatial -omics analysis frameworks and pipeline and reporting limited but promising results with regards to its ability to reproduce results about cellular neighbourhoods, Squidpy does not report its performance in accurately discovering SVGs nor does it quantify how its results relate to those reported in previous studies[28].

## Discussion

Despite being a relatively novel technology, SRT – often alongside scRNA-Seq or other techniques – has already been successfully applied to identify gene expression changes in a variety of tissues and disease states. One example was its application in mouse brains to understand spatially differentially expressed gene involved in early stage Alzheimer's disease[40]. Different SRT methods are best suited to studying different cell types within a tissue to distinguish differences between them in disease states, such as comparing the dopamine neurons from two regions in Parkinson's patients[41]. To further demonstrate how this technology can be applied to an array of conditions and disease, Modlin and colleagues successfully actioned it as part of an investigation into the organisation of cellular subtypes that contribute to the antimicrobial capabilities of human leprosy granulomas[42].

This clear increase in the popularity of SRT has prompted the recent development of many different packages and pipelines for the downstream data analysis of SRT data sets. While it seems that certain studies are still reliant on packages developed for scRNA-Seq data adapted to included SRT analysis such a Seurat[43], the variety of purpose-built available tools will likely replace these. A package for easily identifying SVGs seems to be the most popular aim, and even the pipelines developed so far have centred around this same purpose[20–25,28,33]. However, the scope of developing packages continues to expand to further improve the capabilities of analysis, such as Sparcle, which was developed to be used in conjunction with other packages.

Of all the packages discussed SpatialDE seems to be the most popular, followed by SPARK, Trendsceek and Giotto in terms of being used as benchmarks by which to validate new packages. SpatialDE indicated a tendency to label genes with very low expression as SVGs[22] and certain discrepancies in performance compared to other packages tested on real data sets. This alongside the potential introduction of false positives indicates an area of improvement for this popular package. A current limitation of the validation of package performance is that most commonly two data sets[13], obtained using the same Visium method, are used which will surely introduce inherent bias to the benchmarking process. It would be beneficial to understand the package's performance across datasets from different tissues (instead of exclusively olfactory bulb and breast) generated using a different technology.

To most comprehensively establish the relative performance of all packages, a review should be conducted which benchmarks all packages simultaneously against the same datasets, generated by different SRT methods in different tissues and a standard method for validated established. More packages that are modular and can be integrated alongside one another to expand the scope of analysis are critical and will help advance the field and uptake of this technology. Additionally, the further development of user-friendly pipelines will also make analysing SRT results more accessible. As the array of available tools for analysis of SRT data becomes greater, the results from studies employing the technology will improve and the scope of biological problems that can be addressed will simultaneously expand.

<u>Declarations</u>

**Consent for publication:** All authors provide their consent for publication.

**Competing interest:** The authors declare no competing financial interest.

**Authors' contributions:** NC drafted the manuscript with intellectual input contributions from HTN and MR. HTN and MR reviewed the manuscript. All authors approved the final manuscript.

Figures

**Box 1:**

One key aim of analysing RNA-Seq and scRNA-Seq datasets is to identify differentially expressed genes (DEGs) between two groups from within a group of highly variable genes (HVGs). DEGs are identified between two groups when a gene's expression is statistically significantly different between the two groups is present[1]. While this approach has yielded many important findings, this approach removes organisational context from the groups in question, something that can be recovered using spatial transcriptomics[12]. This new technology has shifted the goal posts for transcriptomics analysis resulting in a sleuth of programmes dedicated to discovering *spatially variable genes* (SVGs)[20–25]. As the name suggests, these genes will have amplified expression in certain regions of the tissue or sample, often displaying an underlying pattern[20,44]. Determining the best method to achieve the most biologically accurate results and computational efficiency is challenging and research in this area is ongoing.

**Figure 1 :** Number of papers when a search was performed using the key words 'Spatial Transcriptomics' using the software 'Publish or Perish'[45] to search PubMed and manually searching bioRvix with the additional parameter of papers published from 01/01/2016-16/04/2021. Papers identified by searching both databases were consolidated, note that this is not a comprehensive view of all papers published on the topic since 2016. Bars in light blue with a dotted outline indicate that not all papers for the calendar year have been included.

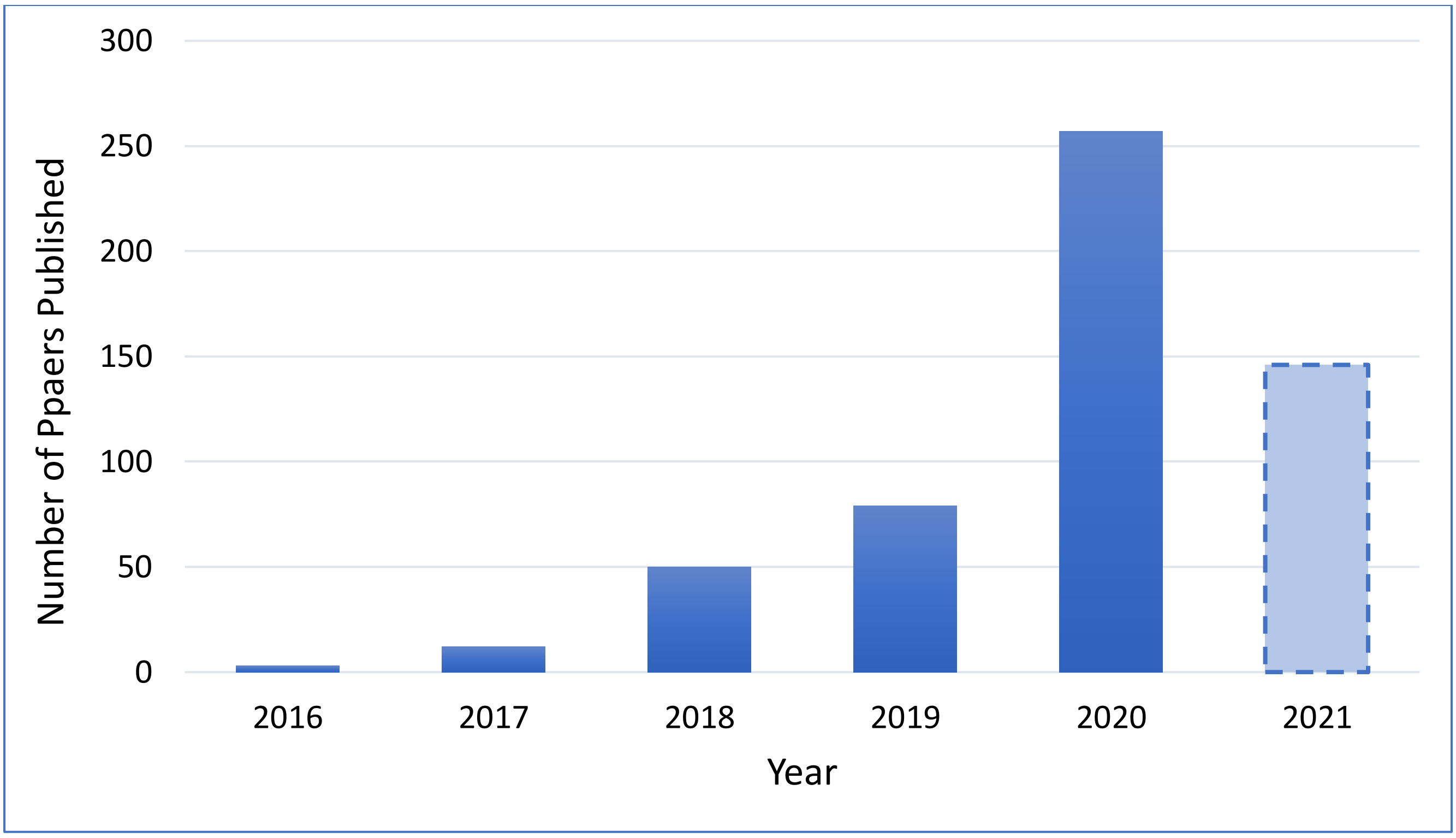

**Table 1:** Comparison of computational packages for analysis of spatial transcriptomics data sets

| Purpose | Package Name | Main Method | Implementation | Pros | Cons | GitHub |
|---|---|---|---|---|---|---|
| **Identifying SVGs** | SpatialDE | GP Regression | Python | Currently most popular package in this category | Labels genes as SVGs that have very low expression & two normalisation steps | https://github.com/Teichlab/SpatialDE |
| | SPARK | Generalised Linear Spatial Model | R | Does not require data to be normalised and controls for type I error | Accuracy not significant improvement on SpatialDE | https://github.com/xzhoulab/SPARK |
| | Trendsceek | Marked Point Process | R | Low false positives reported | Identifies very low number of SVGs & ineffective on larger data sets. | https://github.com/edsgard/trendsceek |
| | BOOST-GP | Bayesian Hierarchical Model | R | Accuracy rate is better than other packages in data sets with many 0 counts | Accuracy rate still low in presence of many 0 counts | https://github.com/Minzhe/BOOST-GP |
| | SOMDE | SOM | Python | Able to efficiently identify SVGs even in very large data sets | In low dropout rate datasets not as good as SpatialDE. | https://github.com/WhirlFirst/somde |
| | scGCO | Graph Cut Algorithm | Python | Results were more reproducible than SpatialDE. Can be used on data sets with millions of cells | ~35% of labelled SVGs were not reproducible | https://github.com/WangPeng-Lab/scGCO |
| **Identifying SVGs + other capabilities** | GP counts | GP Regression | Python | Can determine temporal trajectories and perform pseudotime analysis. | Efficiency on larger data sets unclear | https://github.com/ManchesterBioinference/GPcounts |
| | STUtility | Spatial Autocorrelation | R | Image processing and the ability to create a 3D model from multiple samples. | Accuracy in identifying SVGs and defining tissue heterogeneity not comprehensively reviewed. | https://github.com/jbergenstrahle/STUtility |
| **Assigning lost transcripts** | Sparcle | MLE | Python | Unique capability and can be used alongside other packages. | Developed specifically for smFISH | https://github.com/sandhya212/Sparcle_for_spot_reassignments |
| **Cell type identification** | SpatialDWLS | DWLS | R | *A priori* knowledge can be incorporated | Performance was not validated to other packages on real datasets | https://github.com/rdong08/spatialDWLS_dataset |
| | FICT | Generative Mixture Model | Python | Addresses problem of over-reliance on expression data | Performance drops on data sets with less cells | https://github.com/haotianteng/FICT |
| | RCTD | Supervised Learning | R | Can normalise across platform effects of scRNA-Seq and SRT data sets | Requires well annotated scRNA-Seq data sets | https://github.com/dmcable/RCTD |
| **Spot-to-Spot Clustering** | SpatialCPie | Different Clustering Algorithms | R | Can perform clustering at different resolutions for different subtypes of tissue, cluster graph is novel method of visualising cluster origin in ST | Validation against other packages lacking | https://github.com/jbergenstrahle/SpatialCPie |
| **Pipeline** | Giotto | - | R | Choice of algorithm for identifying marker genes in cell types, dedicated pipelines for lower resolution SRT data | Validation against different biological tissues collected on different platforms lacking. | https://github.com/RubD/Giotto |
| | Squidpy | - | Python | Modular so can incorporate other packages in analysis. | Cellular neighbourhoods not very reproducible. | https://github.com/theislab/squidpy |

**Bibliography:**

1. Exelby, K. *et al.* Precision of Tissue Patterning is Controlled by Dynamical Properties of Gene Regulatory Networks. *Development* **148**, dev.197566 (2021).

2. Hwang, B., Lee, J. H. & Bang, D. Single-cell RNA sequencing technologies and bioinformatics pipelines. *Experimental and Molecular Medicine* vol. 50 96 (2018).

3. Chen, G., Ning, B. & Shi, T. Single-cell RNA-seq technologies and related computational data analysis. *Frontiers in Genetics* vol. 10 (2019).

4. Karaayvaz, M. *et al.* Unravelling subclonal heterogeneity and aggressive disease states in TNBC through single-cell RNA-seq. *Nat. Commun.* **9**, (2018).

5. Regev, A. *et al.* The human cell atlas. *Elife* **6**, (2017).

6. Dong, J. *et al.* Single-cell RNA-seq analysis unveils a prevalent epithelial/mesenchymal hybrid state during mouse organogenesis. *Genome Biol.* **19**, 31 (2018).

7. He, S. *et al.* Single-cell transcriptome profiling of an adult human cell atlas of 15 major organs. *Genome Biol.* **21**, 294 (2020).

8. Ximerakis, M. *et al.* Single-cell transcriptomic profiling of the aging mouse brain. *Nat. Neurosci.* **22**, 1696–1708 (2019).

9. Tiklová, K. *et al.* Single-cell RNA sequencing reveals midbrain dopamine neuron diversity emerging during mouse brain development. *Nat. Commun.* **10**, 1–12 (2019).

10. Zhou, S. *et al.* Single-cell RNA-seq dissects the intratumoral heterogeneity of triple-negative breast cancer based on gene regulatory networks. *Mol. Ther. - Nucleic Acids* **23**, 682–690 (2021).

11. Chen, G., Ning, B. & Shi, T. Single-cell RNA-seq technologies and related

computational data analysis. *Frontiers in Genetics* vol. 10 317 (2019).

12. Marx, V. Method of the Year: spatially resolved transcriptomics. *Nat. Methods* **18**, 9–14 (2021).

13. Ståhl, P. L. *et al.* Visualization and analysis of gene expression in tissue sections by spatial transcriptomics. *Science* vol. 353 78–82 (2016).

14. Lein, E., Borm, L. E. & Linnarsson, S. The promise of spatial transcriptomics for neuroscience in the era of molecular cell typing. *Science* vol. 358 64–69 (2017).

15. Shah, S., Lubeck, E., Zhou, W. & Cai, L. In Situ Transcription Profiling of Single Cells Reveals Spatial Organization of Cells in the Mouse Hippocampus. *Neuron* **92**, 342–357 (2016).

16. Moncada, R. *et al.* Integrating microarray-based spatial transcriptomics and single-cell RNA-seq reveals tissue architecture in pancreatic ductal adenocarcinomas. *Nat. Biotechnol.* **38**, 333–342 (2020).

17. Crosetto, N., Bienko, M. & Van Oudenaarden, A. Spatially resolved transcriptomics and beyond. *Nature Reviews Genetics* vol. 16 57–66 (2015).

18. Asp, M., Bergenstråhle, J. & Lundeberg, J. Spatially Resolved Transcriptomes—Next Generation Tools for Tissue Exploration. *BioEssays* **42**, 1900221 (2020).

19. Waylen, L. N., Nim, H. T., Martelotto, L. G. & Ramialison, M. From whole-mount to single-cell spatial assessment of gene expression in 3D. *Communications Biology* vol. 3 (2020).

20. Svensson, V., Teichmann, S. A. & Stegle, O. SpatialDE: Identification of spatially variable genes. *Nat. Methods* **15**, (2018).

21. Li, Q., Zhang, M., Xie, Y. & Xiao, G. Bayesian modeling of spatial molecular profiling data via Gaussian process. *biorXiv* (2020).

22. Sun, S., Zhu, J. & Zhou, X. Statistical Analysis of Spatial Expression Pattern for Spatially Resolved Transcriptomic Studies. *bioRxiv* (2019) doi:10.1101/810903.

23. Edsgärd, D., Johnsson, P. & Sandberg, R. Identification of spatial expression trends in single-cell gene expression data. *Nat. Methods* **15**, (2018).

24. Hao, M., Hua, K. & Zhang, X. SOMDE: A scalable method for identifying spatially variable genes with self-organizing map. *bioRxiv* (2021) doi:https://doi.org/10.1101/2020.12.10.419549.

25. Zhang, K., Feng, W. & Wang, P. Identification of spatially variable genes with graph cuts. *bioRxiv* 491472 (2018) doi:10.1101/491472.

26. BinTayyash, N. *et al.* Non-parametric modelling of temporal and spatial counts data from RNA-seq experiments. *bioRxiv* 2020.07.29.227207 (2020) doi:10.1101/2020.07.29.227207.

27. Bergenstråhle, J., Larsson, L. & Lundeberg, J. Seamless integration of image and molecular analysis for spatial transcriptomics workflows. *BMC Genomics* **21**, 482 (2020).

28. Palla, G. *et al.* Squidpy: a scalable framework for spatial single cell 2 analysis. *bioRxiv* 2021.02.19.431994 (2021) doi:10.1101/2021.02.19.431994.

29. Prabhakaran, S., Nawy, T. & Pe'er', D. Sparcle: assigning transcripts to cells in multiplexed images 1. *bioRxiv* 2021.02.13.431099 (2021) doi:10.1101/2021.02.13.431099.

30. Levine, J. H. *et al.* Data-Driven Phenotypic Dissection of AML Reveals Progenitor-like

Cells that Correlate with Prognosis. *Cell* **162**, 184–197 (2015).

31. Neal, R. M. *Markov Chain Sampling Methods for Dirichlet Process Mixture Models*. *Journal of Computational and Graphical Statistics* vol. 9 http://www.jstor.org/about/terms.html. (2000).

32. Dong, R. & Yuan, G.-C. SpatialDWLS: accurate deconvolution of spatial transcriptomic data. *bioRxiv* 2021.02.02.429429 (2021) doi:10.1101/2021.02.02.429429.

33. Dries, R. *et al.* Giotto, a pipeline for integrative analysis and visualization of single-cell spatial transcriptomic data. *bioRxiv* (2019) doi:10.1101/701680.

34. Teng, H., Yuan, Y. & Bar-Joseph, Z. Cell Type Assignments for Spatial Transcriptomics Data. *bioRxiv* 2021.02.25.432887 (2021) doi:10.1101/2021.02.25.432887.

35. Cable, D. M. *et al.* Robust decomposition of cell type mixtures in spatial transcriptomics. *bioRxiv* 2020.05.07.082750 (2020) doi:10.1101/2020.05.07.082750.

36. Bergenstråhle, J., Bergenstråhle, L. & Lundeberg, J. SpatialCPie: An R/Bioconductor package for spatial transcriptomics cluster evaluation. *BMC Bioinformatics* **21**, 161 (2020).

37. Lun, A. T. L., McCarthy, D. J. & Marioni, J. C. A step-by-step workflow for low-level analysis of single-cell RNA-seq data with Bioconductor. *F1000Research* **5**, (2016).

38. Finak, G. *et al.* MAST: A flexible statistical framework for assessing transcriptional changes and characterizing heterogeneity in single-cell RNA sequencing data. *Genome Biol.* **16**, 278 (2015).

39. Jiang, L., Chen, H., Pinello, L. & Yuan, G. C. GiniClust: Detecting rare cell types from single-cell gene expression data with Gini index. *Genome Biol.* **17**, 144 (2016).

40. Navarro, J. F. *et al.* Spatial Transcriptomics Reveals Genes Associated with Dysregulated Mitochondrial Functions and Stress Signaling in Alzheimer Disease. *iScience* **23**, (2020).

41. Aguila, J. *et al.* Spatial transcriptomics identifies novel markers of vulnerable and resistant midbrain dopamine neurons. *bioRxiv* (2018) doi:10.1101/334417.

42. Ma, F. *et al.* Single Cell and Spatial Transcriptomics Defines the Cellular Architecture of the Antimicrobial Response Network in Human Leprosy Granulomas. *bioRxiv* 2020.12.01.406819 (2020) doi:10.1101/2020.12.01.406819.

43. Ortiz, C. *et al.* Molecular Atlas of the Adult Mouse Brain. *bioRxiv* (2019) doi:10.1101/784181.

44. Hu, J. *et al.* Integrating gene expression, spatial location and histology to identify spatial 1 domains and spatially variable genes by graph convolutional network 2 3. *bioRxiv* 2020.11.30.405118 (2020) doi:10.21203/RS.3.RS-119776/V1.

45. Harzing, A.-W. Publish or perish? *Harzing.com* https://harzing.com/resources/publish-or-perish/os-x (2016). Access date: 26/04/2021.